\documentclass[10pt,letterpaper,journal]{IEEEtran}
\usepackage{cite}
\usepackage{mathtools, cuted}
\usepackage[pdftex]{graphicx}
\usepackage{placeins}
\usepackage[thmmarks]{ntheorem}
\usepackage{xfrac}
\usepackage{times}
\usepackage{amsmath}
\usepackage[capitalize]{cleveref}

\usepackage{booktabs}
\usepackage{tabularx}
\usepackage{color}
\usepackage{colortbl}
\usepackage{xcolor,colortbl}
\usepackage{float}
\usepackage[normalem]{ulem}
\usepackage[T1]{fontenc}
\usepackage{array,booktabs}
\usepackage{url}
\usepackage{amssymb}
\usepackage{subfigure}
\usepackage{tabu}
   \usepackage{booktabs}
\usepackage{multirow}
\usepackage{multicol}
\usepackage{hhline}

\newcolumntype{P}[1]{>{\centering\arraybackslash}p{#1}}
\newcolumntype{M}[1]{>{\centering\arraybackslash}m{#1}}

\def \ggr {\textit{PAR}_\text{r}}

\def \thetaz {\theta_\text{z}}
\def \gammas {\gamma_\text{s}}
\def \Iglobal {\textit{I}_\text{global}}
\def \Idiff {\textit{I}_\text{diff}}

\def \thetaF {\theta_{\text{F}} }

\def \hs {\textit{h}_\text{s} }
\def \ls {\textit{l}_\text{s} }
\def \lsbot {\textit{l}_{\text{s} \lvert \text{b} }}
\def \lstop {\textit{l}_{\text{s} \lvert \text{t} }}

\def \RA {\textit{R}_\text{A}}

\def \etadiffF {\eta_{\text{diff} \lvert \text{F}}}
\def \etadiffB {\eta_{\text{diff} \lvert \text{B}}}

\def \thetaone {\theta_\text{1}}

\def \thetatwo {\theta_\text{2}}

\def \baseL { \frac{\textit E + \textit h \cdot \text{sin}\beta }{\text{tan}\beta}  }

\def \baseS  { \frac{\textit E}{\text{tan}\beta}  }

\def \Ignddiff {\textit{I}_\text{gnd;diff}}
\def \Ignddir {\textit{I}_\text{gnd;dir}}

\def \FdzgndF {\textit{F}_{\text{d}\textit{z} \rightarrow \text{gnd} \lvert \text{F}}}
\def \FdzgndB {\textit{F}_{\text{d}\textit{z} \rightarrow \text{gnd} \lvert \text{B}}}

\def \IPVAlbdiffF {\textit{I}_{\text{M,Alb:diff} \lvert \text{F}}}
\def \IPVAlbdiffB {\textit{I}_{\text{M,Alb:diff} \lvert \text{B}}}
\def \IPVAlbdiff {\textit{I}_{\text{M,Alb:diff}}}

\def \psitF {\psi_{ \text {t} \lvert \text{F}}^\textit{(i)}}
\def \psibF {\psi_{ \text {b} \lvert \text{F}}^\textit{(i)}}
\def \psitB {\psi_{ \text {t} \lvert \text{B}}^\textit{(i)}}
\def \psibB {\psi_{ \text {b} \lvert \text{B}}^\textit{(i)}}

\def \psitFF {\psi_{ \text {t} \lvert \text{F}}^{(\textit{i} + \text{1})}}
\def \thetaFt{ {\theta_\text{t|F}}^\textit{(i)}}
\def \thetaFb {{\theta_\text{b|F}}^\textit{(i)}}
\def \thetaBt {{\theta_\text{t|B}}^\textit{(i)}}
\def \thetaBb {{\theta_\text{b|B}}^\textit{(i)}}
\def \psibFF {\psi_{ \text {b} \lvert \text{F}}^{(\textit{i} + \text{1})}}

\def \FdzUgndF {\textit{F}_{\text{d}\textit{z} \rightarrow \text{Ugnd} \lvert \text{F}}}
\def \FdzUgndB {\textit{F}_{\text{d}\textit{z} \rightarrow \text{Ugnd} \lvert \text{B}}}

\def \IPVAlbdirF {\textit{I}_{\text{M,Alb:dir} \lvert \text{F}}}
\def \IPVAlbdirB {\textit{I}_{\text{M,Alb:dir} \lvert \text{B}}}
\def \IPVAlbdir {\textit{I}_{\text{M,Alb:dir}}}

\def \yeav {\textit{Y}_\text{e} \text{(AV)}}
\def \yepv {\textit{Y}_\text{e} \text{(PV)}}

\begin{document}

\title{Module Technology for Agrivoltaics: Vertical Bifacial vs. Tilted Monofacial Farms}

\author{Muhammad~Hussnain~Riaz,~Rehan~Younas,~Hassan~Imran,~\IEEEmembership{Student~Member,~IEEE,} Muhammad~Ashraful~Alam,~\IEEEmembership{Fellow,~IEEE,}~and~Nauman~Zafar~Butt,~\IEEEmembership{Member,~IEEE}

\thanks{This manuscript is submitted to IEEE Journal of Photovoltaics.}
\thanks{Muhammad Hussnain Riaz, Rehan Younas, Hassan Imran, and Nauman Zafar Butt are with Department of Electrical Engineering, School of Science and Engineering, Lahore University of Management Sciences, Lahore 54792, Pakistan. (hussnainriaz8@gmail.com;
reh.younas@gmail.com;
hassan.imran.ee@gmail.com; nauman.butt@lums.edu.pk) Muhammad Ashraful Alam is with School of Electrical and Computer Engineering, Purdue University, West Lafayette, IN 47907, USA. (e-mail: alam@purdue.edu)}
} 

\maketitle


\begin{abstract}
Agrivoltaics is an innovative approach in which solar photovoltaic (PV) energy generation is collocated with agricultural production to enable food-energy-water synergies and landscape ecological conservation. This dual-use requirement leads to unique co-optimization challenges (\textit{e.g.} shading, soiling, spacing) that make module technology and farm topology choices distinctly different from traditional solar farms. Here we compare the performance of the traditional optimally-titled North/South (\textit{N/S}) faced monofacial farms with a potential alternative based on vertical East/West (\textit{E/W})-faced bifacial farms. Remarkably, the vertical farm produces essentially the same energy output and photosythetically active radiation (\textit{PAR}) compared to traditional farms as long as the PV array density is reduced to half or lower relative to that for the standard ground mounted PV farms. Our results explain the relative merits of the traditional monofacial vs. vertical bifacial farms as a function of array density, acceptable \textit{PAR}-deficit, and energy production. The combined \textit{PAR}/Energy yields for the vertical bifacial farm may not always be superior, it could still be an attractive choice for agrivoltaics due to its distinct advantages such as minimum land coverage, least hindrance to the farm machinery and rainfall, inherent resilience to PV soiling, easier cleaning and cost advantages due to potentially reduced elevation..

\end{abstract}

\section{Introduction}
\label{Introduction}
While the concept of Agrivoltaics (AV) dates back to 1980s \cite{goetzberger1982coexistence}, the dramatic reduction of solar modules over the last decade and world-wide proliferation of PV technologies have made the approach potentially viable in many parts of the world. This dual approach of harvesting energy and food together in a given land area can maximize the land productivity with additional synergistic benefits including reduced water budget, improved crop yield, agricultural land preservation, and, socio-economic welfare of farmers \cite{dinesh2016potential,weselek2019agrophotovoltaic,mead1980concept,elamri2018water,genccer2017directing,parkinson2020economic,schindele2020implementation,adeh2018remarkable}. AV farming can offer an attractive solution to a potential conflict between rapidly spreading ground mounted solar photovoltaic parks and the agricultural production, landscape ecology and biodiversity \cite{majumdar2018dual, barron2019agrivoltaics}. Moreover, AV can be leveraged to enable crop resilience against the increasing climate change vulnerabilities, such as the excessive heat stress and drought, in particular for hot and arid climates~\cite{elamri2018water}. Other synergistic benefits may include sharing of water between the cleaning of panels and irrigation to reduce the overall water and maintenance cost.

The AV farm design has so far explored the traditional fixed tilt \textit{N/S} faced PV arrays or solar tracking schemes with limited focus on comparative analysis between module technologies. The fixed tilt systems have primarily been explored in two configurations: full and half PV array densities that correspond to row to row pitch being twice and four times the height of the panels, respectively. Durpraz \cite{dupraz2011combining, dupraz2011mix} predicted that 35-73$\%$ increase in land productivity was possible for Montpelier ($\text{43}^\circ$N), France, when solar arrays were arranged at full and half densities respectively. Marrou~\cite{marrou2013microclimate} performed AV experiments in Montpelier, France using full and half density arrays for both short cycle (lettuce, cucumber) and long cycle (durum wheat) crops. The study emphasized that the main focus for developing AV system should be on exploring mitigation strategies for light reduction and optimal selection of crops~\cite{barron2019agrivoltaics}. Majumdar~\cite{majumdar2018dual} modeled AV system at Pheonix Arizona and showed that south faced solar arrays titled at 30$^\circ$ received 60$\%$ and 80$\%$ of the total global radiation for half and quarter density of solar arrays respectively as compared to an open field. Malu~\cite{malu2017agrivoltaic} modeled AV grape farming for Maharashtra (19.59$^\circ$ N), India, and predicted that it could increase the economic value of the land by 15 times as compared to the conventional farming.

Amaducci~\cite{amaducci2018agrivoltaic} showed that the reduction of global radiation under \textit{AV} was affected more by panel density than by dynamic management of tilt through solar tracking. Sekiyama~\cite{sekiyama2019solar} did an experiment in Ichihara City (35.37$^\circ$ N, 140.13$^\circ$ W), Japan, using PV arrays at full vs. half density and found that the corn yield for the half density remarkably outperformed that of the open (control) farm by 5.6\%. 

Although the reported studies have successfully demonstrated the potential of \textit{AV} across various climates and crops, the choice of module technology and farm topology have not been explored. Technology innovations to address practical challenges including the need to elevate the panels to 10-15 \textit{ft}, the issue about soiling and the difficulty to clean at high elevation, \textit{etc.} have so far not been considered. Among the available module technologies, the potential for bifacial solar panels has not yet been studied in detail although the bifacial PV is attracting a great attraction in the commercial PV market due to higher performance and temperature insensitivity\cite{patel2020temperature}. In this paper, we address these issues by exploring the vertically tilted \textit{E/W} faced bifacial (\textit{bi-E/W}) panels, which can be of great potential interest for AV due to a number of practical advantages including inherent resilience to soiling losses, minimum land coverage that provides least hindrance to the farm machinery and rainfall, and possibility to mount much closer to the ground hence offering significantly lower installation cost and easier access for cleaning \cite{guo2013vertically}\cite{chudinzow2020vertical}.The soiling losses can be substantial for tilted panels in many climates in south Asia, middle East, and north Africa, where power losses due to soling in tilted panels could be as high as ~1$\%$ per day \cite{ullah2019investigation}\cite{hegazy2001effect}\cite{al2013design}. Soiling losses are known to mitigate strongly as the panel tilt varies from horizontal to vertical \cite{ullah2019investigation}. For this reason, in arid climates where long dry periods (between successive rain events) can exacerbate PV soiling losses and drought concerns for the crops, vertical panel AV arrangement could be especially advantageous. Despite of these potential attractions, there is no study currently available in literature to explore relative potential for the vertical bifacial AV farming.

In this paper, we present a computational investigation into the vertical bifacial AV farms that for the first time explores its relative advantages and tradeoffs as compared to the conventional \textit{N/S} tilted monofacial and bifacial AV farms. By developing a rigorous model for the radiation interception at the panels and \textit{PAR} transmission to the crops respectively, we specifically address the following questions when comparing these AV orientations: (i) What are the relative food-energy implications as the PV array density is varied? (ii) How does the tilt angle change the relative food-energy production? and (iii) What land productivities can be achieved relative to the standalone energy and crop systems across the year? The primary goal here is to establish the comparative performance of vertical AV farms using traditional metrics i.e., PV energy and PAR without considering the complex crop modeling which is beyond the scope of this paper. This approach may be imperfect in terms of evaluating the specific crop yields, but would offer guidance and establish principles of technology assessment.

This paper is divided into four section. In \cref{Methodology} we describe the detailed methodology. Results are discussed in \cref{results} whereas conclusions are furnished in \cref{conclusions}.

\section{Methodology}
\label{Methodology}
\begin{figure}[!b]
\centering
\subfigure
  {
  \includegraphics[width=1\linewidth]{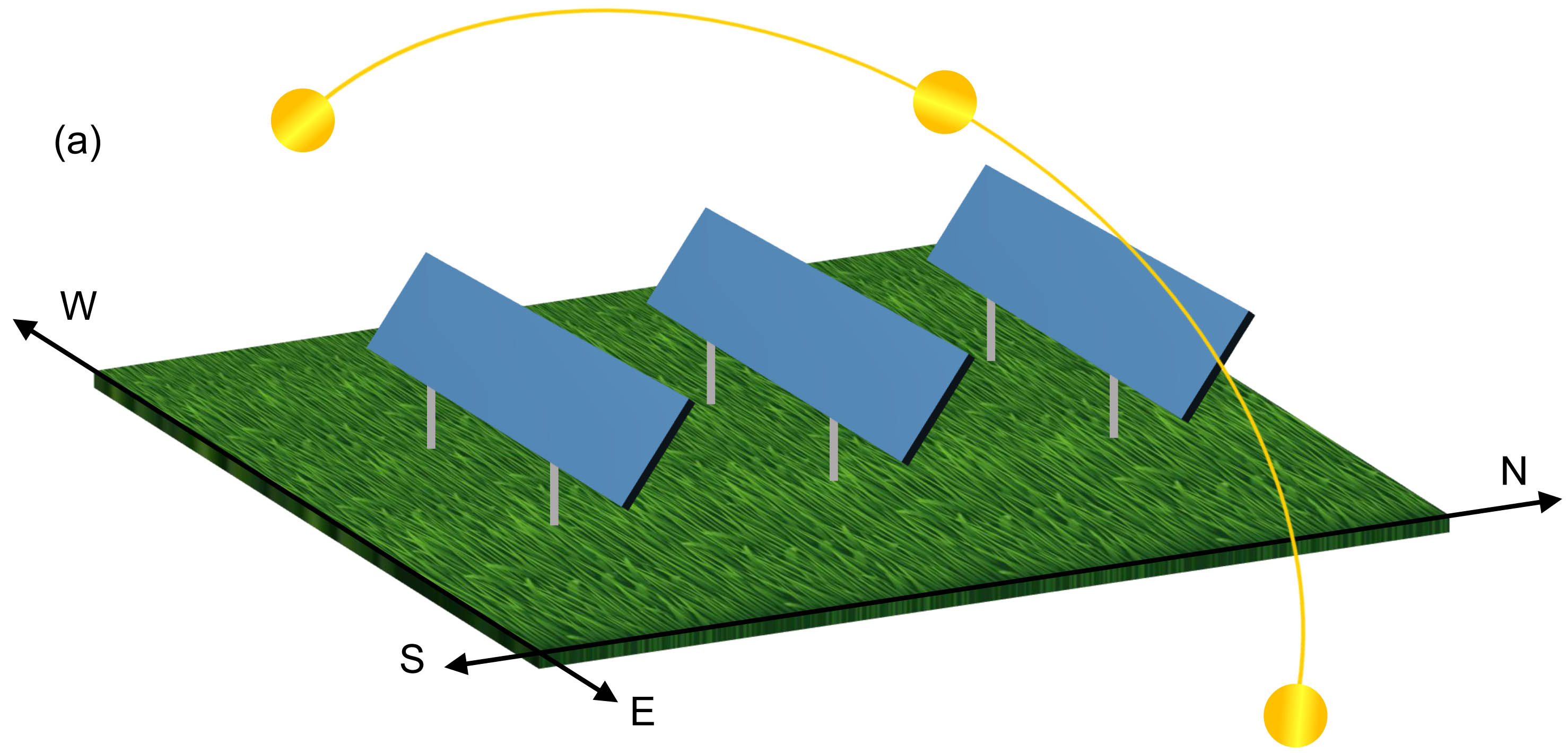} 
  }
\subfigure
{
  \includegraphics[width=1\linewidth]{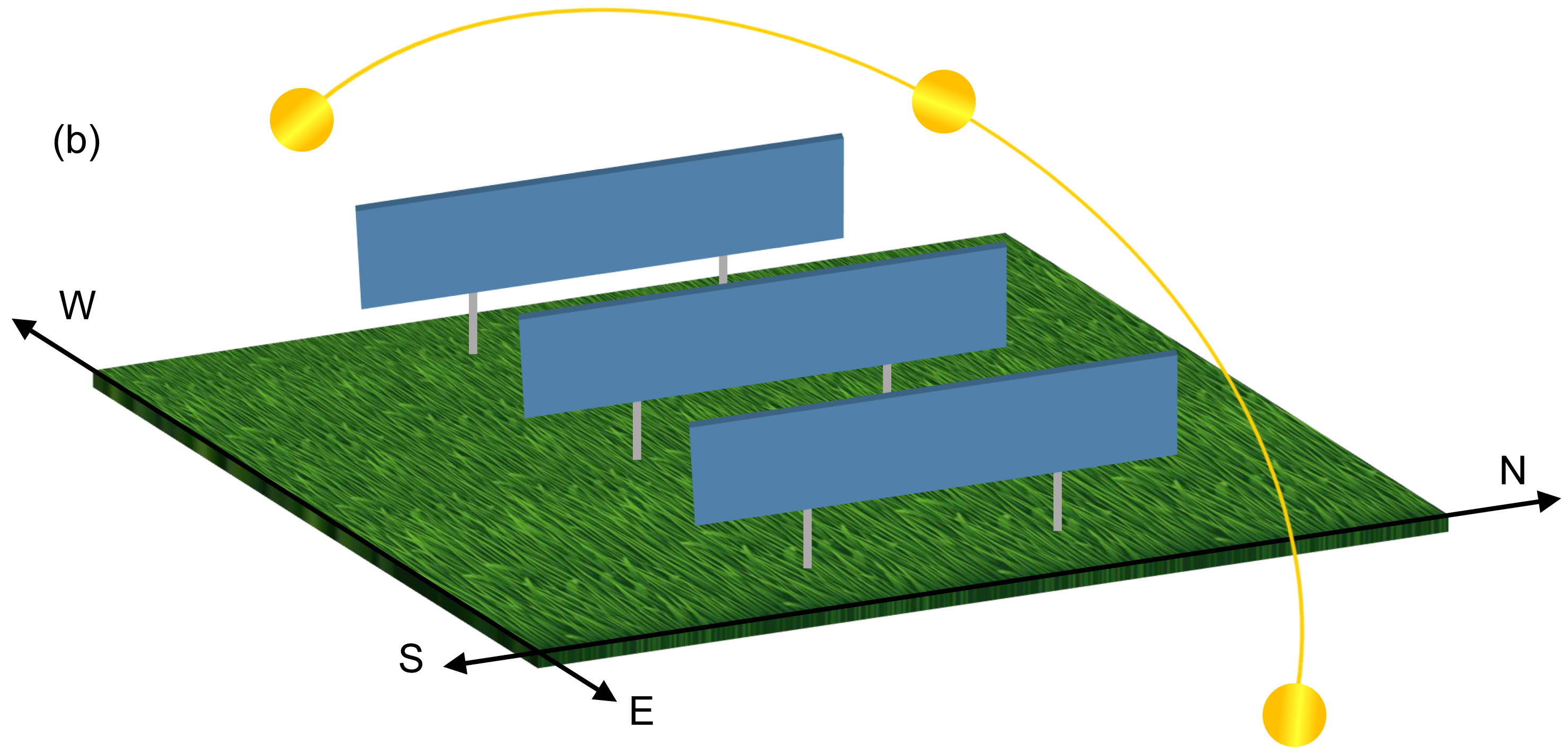} 
}
\caption{(a) Tilted monofacial south facing (\textit{mono-N/S}), and (b) vertical bifacial east-west facing (\textit{bi-E/W}) solar modules.}
\label{Fig_1}
\end{figure}

For calculating energy harvested by solar modules, we assume that arrays are long enough so that the edge effects are negligible. This allows for solving the PV energy generation in two-dimensions assuming symmetry in the third (infinitely long) dimension parallel to the row. The solar farm consists of solar modules of height (\textit{h}) that are mounted at an elevation (\textit{E}) above the ground and the pitch (row-to-row separation) between them is \textit{p}. The two PV orientations, i.e., \textit{N/S} and \textit{E/W} are illustrated in Fig.\ref{Fig_1}. 

\subsection{Solar irradiation and PV Models}
Recent work presented in \cite{patel2019worldwide} provides a detailed modeling approach for solar energy harvested in ground mounted bifacial PV arrays including the effects of self/mutual shading. For AV applications, however, PV arrays typically need to be elevated above ground to conveniently allow agricultural operations hence necessitating models that work for elevated PV structures. To precisely calculate the shading patterns for the crops and the PV energy generation for elevated panels, we develop new analytical models to calculate the direct/diffused albedo collection by the elevated PV array and the diffused \textit{PAR} transmitted under the panels including the masking by the elevated PV array. Model validation is done by comparison to published experimental data (\textit{Section S1}). 

\subsubsection{Calculation of Global Horizontal, Direct and Diffuse Irradiance}

The global horizontal irradiance ($\Iglobal$) and its components, \textit{i.e.}, direct normal (\textit{Idir}) and diffused horizontal irradiance (\textit{Idiff}) along with the sun’s trajectory (defined by zenith ($\thetaz$) and azimuth ($\gammas$) angles) at any location (latitude, longitude) were calculated using NREL’s algorithm ~\cite{reda2004solar} implemented in Sandia’s photovoltaic modeling library (PVLib) \cite{pvlibnew}. Huarwitz clear sky model \cite{haurwitz1945insolation} was used to calculate Iglobal with a one-minute time resolution. Satellite derived data from NASA Surface Meteorology and Solar Energy database \cite{nasalarc} was used to calibrate the location-specific climate factors similar to the approach used in \cite{khan2017vertical}\cite{hussnain2020PVSC}\cite{hasan2020PVSC}.

\begin{figure}[!b]
    \centering
    \includegraphics[width=1.0\linewidth]{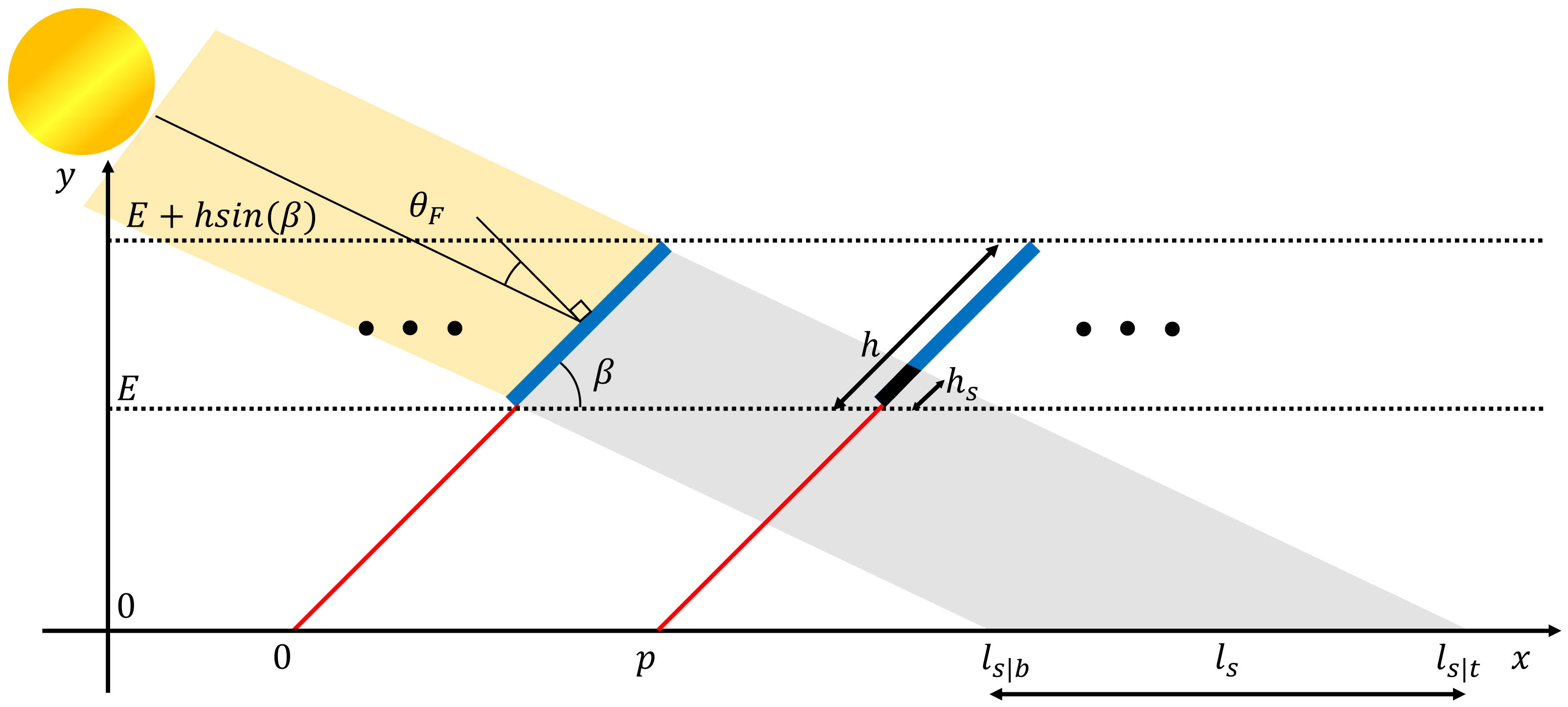}
    \caption{Illustration of direct sunlight incident on solar modules of height \textit{h}, tilted at an angle $\beta$ and mounted at an elevation \textit{E} above the ground. The shadow length $\textit{l}_\text{s}$ on the ground and shadow height on the adjacent module $\textit{h}_\text{s}$ along with angle of incident $\thetaF$ are also depicted.} 
    \label{Fig_2}
\end{figure}

\subsubsection{Calculation of Shadow Lengths}
The shadow for the direct beam of the sun along the pitch is calculated on the ground or at any elevation below the PV array. The shadow length at the ground due to the top and bottom points of the module (denoted by $\lstop$  and $\lsbot$   respectively, see Fig.\ref{Fig_2}) is given as:

\begin{equation}
  \renewcommand{\arraystretch}{1.2}
  \left.
\begin{array}
{r@{\;}l}
{\lstop = \frac{(\textit{E} + h\cdot \text{sin}\beta )\cdot \text{cos}\thetaF}{\text{sin}(90-\beta + \thetaF)\cdot \text{sin}\beta } } \\
{\lsbot = \frac{\textit{E}\cdot \text{cos}\thetaF}{\text{sin}(90-\beta + \thetaF)\cdot \text{sin}\beta }}
\end{array}
\right\}
\label{eq_5}
\end{equation}

where $\thetaF$ is the angle of incidence \cite{patel2019worldwide} at the front surface and $\beta$ is the tilt angle. The actual shadow length (\textit{ls}) on the ground is given by $\ls=\lstop- \lsbot$. If $\ls>p$, then a part of the shadow with height denoted by $\hs$ will be dropped on the adjacent module (see Fig. \ref{Fig_2}) causing the mutual shading:

\begin{equation}
\hs = \frac{\text{sin}\left [ \text{90} + \thetaF-\beta \right ]\times \left [\lstop - \textit p \right ]}{\text{sin}\left [ \text{90} -\thetaF \right ]} - \frac{\textit E }{\text{sin}(\beta)}
\label{eq_6}
\end{equation}
The mutual shading between modules becomes more significant as \textit{p/h} is reduced.

\subsubsection{Energy Harvested from Direct and Diffuse Irradiance}
The calculation for the direct and diffused radiation intercepted by the elevated panels is identical to that for ground mounted panels so the existing modeling approaches described in \cite{patel2019worldwide}\cite{khan2017vertical} are applied. The efficiency of the front/back surfaces of the module under direct and diffused light are taken as 19$\%$ and 16$\%$ respectively.

\subsubsection{Energy Harvested From Direct Albedo Irradiance}

\begin{figure}[!t]
    \centering
    \includegraphics[width=1.0\linewidth]{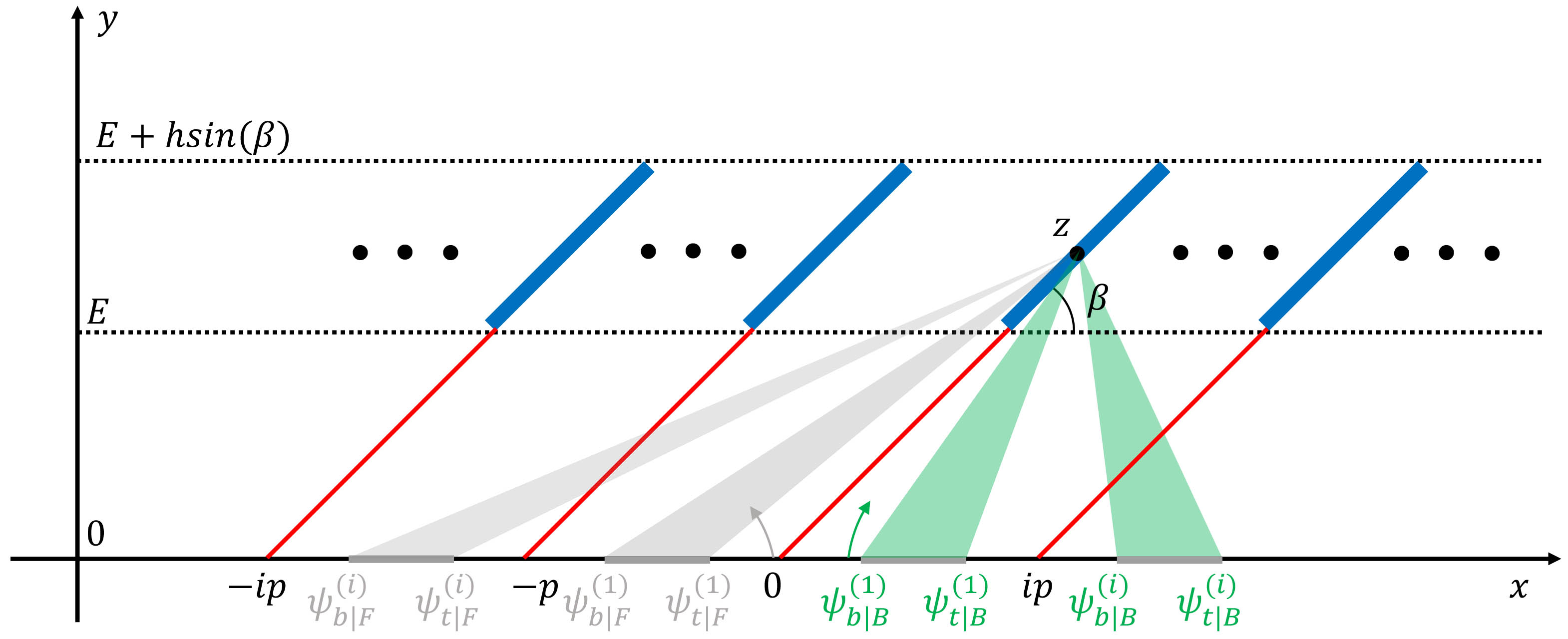}
    \caption{The angles subtended by the edges of the shade on the ground due to direct irradiance at a given time on the front and back surfaces of the modules. The green and gray arrow represent that angles are measured from the ground in clockwise and anti-clockwise direction respectively.}
    \label{Fig_3}
\end{figure}

To compute the collection of direct sunlight albedo at any point z on the module, the angles $\psitF$,$\psibF$  ,$\psitB$  and $\psibB$ subtended by the edges of the shadow (\textit{i.e.} $\lstop$ and $\lsbot$) from the ground to the front and back faces of the module in the $i^{th}$ pitch respectively are calculated (see \cref{Fig_3}) and are given by: 
\begin{equation}
  \renewcommand{\arraystretch}{1.2}
  \left.
\begin{array}
{r@{\;}l}
{\psitF(z) = \text{tan}^{- \text 1} \Big [ \frac{ \textit E + \textit z \cdot \text{sin}\beta }{ i \times p - \lstop + \frac{E + z\cdot \text{sin}\beta}{\text{tan}\beta}} \Big ] } \\
{\psibF(z) = \text{tan}^{- \text 1} \Big [ \frac{ \textit E + \textit z \cdot \text{sin}\beta }{ i \times p - \lsbot + \frac{E + z\cdot \text{sin}\beta}{\text{tan}\beta}} \Big ] } \\
{\psitB(z) = 180^\circ - \text{tan}^{- \text 1} \Big [ \frac{ \textit E + \textit z \cdot \text{sin}\beta }{ -(i-1) \times p - \lstop + \frac{E + z\cdot \text{sin}\beta}{\text{tan}\beta}} \Big ] } \\
{\psibB(z) = 180^\circ - \text{tan}^{- \text 1} \Big [ \frac{ E + z \cdot \text{sin}\beta }{ -(i-1) \times p - \lsbot + \frac{E + z\cdot \text{sin}\beta}{\text{tan}\beta}} \Big ] } 
\end{array}
\right\}
\label{eq_13}
\end{equation}
The view factors for front surface of the module from point \textit{z} to the unshaded part of the ground are given as:

\begin{equation}
\FdzUgndF (z) = \frac{\text 1}{\text 2} \times
\begin{cases}
\sum \limits_{\textit{i}} \left \{ \text{sin} \left ( \psibF \right )
- \text{sin} \left ( \psitFF \right ) \right \} + \\
\text{sin}\beta - \text{sin}\left (\psitF \right ) \,\,\,\,\, {;~\text{if}~\thetaF \le \text{90}^\circ} \\
\sum \limits_{\textit{i}}
\left \{ \text{sin} \left ( \psitF \right )
- \text{sin} \left ( \psibFF \right ) \right \} + \\
\text{sin}\beta - \text{sin}\left (\psibF \right ) \,\,\,\,\, {;~\text{if}~\thetaF > \text{90}^\circ}
\end{cases}
\label{eq_14}
\end{equation}
Similarly for the back surface, $\FdzUgndB$ can also be computed based upon eq. (4).

The power received by at any point $\textit{z}$ on the front and back faces of module due to direct albedo is given by
\begin{equation}
  \renewcommand{\arraystretch}{1.2}
  \left.
\begin{array}
{r@{\;}l}
{\IPVAlbdirF (z) = \Ignddir \times \etadiffF \times \RA \times \FdzUgndF(\textit{z})   } \\
{\IPVAlbdirB (z) = \Ignddir \times \etadiffB \times \RA \times \FdzUgndB(\textit{z}) } 
\end{array}
\right\}
\label{eq_16}
\end{equation}
where $\RA$ is the ground albedo and considered equal to 0.25 in all calculations. The total power ($\IPVAlbdir$)  generated by direct albedo per unit solar farm area is given as

\begin{equation}
\IPVAlbdir = \frac{1}{p} \times \int_\text{0}^\textit{h} \left [ \IPVAlbdirF(z) + \IPVAlbdirB(z) \right ] \textit{dz} 
\label{eq_17}
\end{equation}

\subsubsection{Energy Harvested From Diffuse Irradiance}

\begin{figure}[!b]
    \centering
    \includegraphics[width=1.0\linewidth]{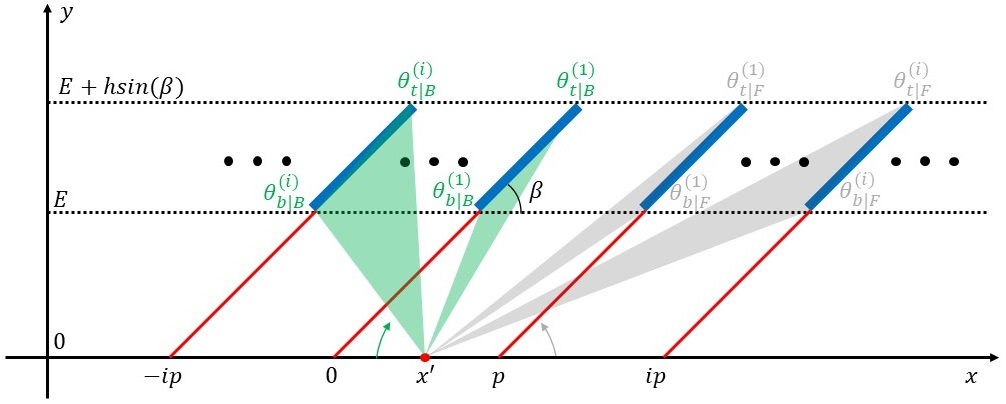}
    \caption{Masking of diffuse albedo light on the face of the module by the adjacent modules in either direction.}
    \label{Fig_4}
\end{figure}
For any point \textit{x} between the two adjacent panels $\left [ \text 0 \le \textit{x} \le \textit{p} \right ]$, the masking angles ($\thetaFt$ and $\thetaFb$) for the diffused sunlight subtended from $\textit{x}$ to the top and bottom points of PV modules at the back surfaces (see \cref{Fig_4}) are given by:

\begin{equation}
  \renewcommand{\arraystretch}{1.2}
  \left.
\begin{array}
{r@{\;}l}
{\thetaBt (x) = \text{180} - \text{tan}^{-\text1} \bigg [ \frac{ E+h\cdot \text{sin}\beta} {(i - \text 1)p - x + \baseL} \bigg ] } \\
{\thetaBb (x) = \text{180} - \text{tan}^{-\text1} \bigg [ \frac{ E} {(i- \text 1)p - x + \baseS} \bigg ]}
\end{array}
\right\}
\label{eq_ch8_11}
\end{equation}
The total masking angle for $\textit{x}$ for the back side of the panel over the entire farm is given as
\begin{equation}
\Delta \thetaone = \int_0^p \bigg [ \sum_i \Big ( \thetaBt - \thetaBb \Big )  \bigg] dx
\label{eq_ch8_12}
\end{equation}
Similarly, masking angles for front side of the panel \textit{i.e.} $\thetaFt$ and $\thetaFb$ are given as
\begin{equation}
  \renewcommand{\arraystretch}{1.2}
  \left.
\begin{array}
{r@{\;}l}
{\thetaFt (x) = \text{tan}^{-\text1} \bigg [ \frac{ E+h\cdot \text{sin}\beta} {(i- \text 1)p -x + \baseL} \bigg ] } \\
{\thetaFb (x) = \text{tan}^{-\text1} \bigg [ \frac{ E} {(i- \text 1)p - x + \baseS} \bigg ]}
\end{array}
\right\}
\label{eq_ch8_13}
\end{equation}
The total masking angle for $\textit{x}$ for the front side of the panel  over the entire farm is given as
\begin{equation}
\Delta \thetatwo = \int_0^p \bigg [ \sum_i \Big ( \thetaFt - \thetaFb \Big )  \bigg] dx
\label{eq_ch8_14}
\end{equation}
The average diffused light reaching the ground is given as \cite{khan2017vertical}
\begin{equation}
\Ignddiff = \Idiff \times \frac{\text 1}{\text 2} \Big [ \text{cos}(\Delta \thetaone) + \text{cos}(\Delta \thetatwo)  \Big]
\label{eq_ch8_15}    
\end{equation}

The power generated by diffuse albedo for any point $z$ by the front and back surfaces of the module is then given by

\begin{equation}
  \renewcommand{\arraystretch}{1.2}
  \left.
\begin{array}
{r@{\;}l}
{\IPVAlbdiffF(z) =  \etadiffF \times \Ignddiff \times \textit{R}_\text{A} \times \FdzgndF } \\
{\IPVAlbdiffB(z) =  \etadiffB \times \Ignddiff \times \textit{R}_\text{A} \times \FdzgndB}
\end{array}
\right\}
\label{eq_20}
\end{equation}

where $\FdzgndF$  and $\FdzgndB$  are view factors at any point z along the module height masked by the front and back surfaces of the module to the ground respectively \cite{patel2019worldwide}. 
The total power generated ($\IPVAlbdiff$) by diffused albedo per unit solar farm area is given as:

\begin{equation}
\IPVAlbdiff = \frac{1}{p} \times \int_\text{0}^\textit{h} \left [ \IPVAlbdiffF(z) + \IPVAlbdiffB(z) \right ] \textit{dz} 
\label{eq_21}
\end{equation}

The total PV power generated ($\textit{I}_\text{M}$) per unit farm area is then calculated as:

\begin{equation}
\textit{I}_\text{M} = \textit{I}_\text{M,dir} + \textit{I}_\text{M,diff} + \IPVAlbdir + \IPVAlbdiff
\label{eq_212}
\end{equation}

\subsection{The transmitted PAR under the panels}
At any given time, the observation point (OP) under the panels along the pitch is either under shade hence receiving the diffused light only, $\textit{I}_\text{op} = \Ignddiff$ or receiving both direct as well as diffused sunlight ($\textit{I}_\text{op} = \Ignddir + \Ignddiff$). The relative cumulative radiation ($PAR_r$) incident at a specific crop height is calculated as the percentage ratio of light received under panel coverage to the total incident light with no panels installed:
\begin{equation}
\ggr(\%) = \frac{\text{1}}{\textit{k}} \sum_\text{op=1}^\textit{k} \frac{\textit{I}_\text{op}}{\Iglobal} \times \text{100\%}
\label{eq_24}    
\end{equation}

Here, $\textit{k}$ is varied to cover the entire region with in a pitch. 

\subsection{Calculation of Land Productivity Factor}
Typically, land equivalent ratio \textit{LER}, a metric that was originally defined for measuring efficacy of inter-cropping \cite{mead1980concept}, has been used to characterize the productivity of land for AV relative to the standalone PV systems and agricultural farms. It however requires the estimation of the relative crop yields which may have a wide variation across different crop species and can have a complex dependence on the dynamically varying shades and other microclimate factors which are beyond the scope of the present work. Here, for the purpose of comparing various module technologies and farm topologies, we define a simpler metric named land productivity factor (\textit{LPF}) a crop independent parameter that estimates the AV land productivity by adding the relative yields for PV energy and the transmitted PAR at the crop level:

\begin{equation}
\textit{LPF} = \frac{\ggr}{100} + \frac{\yeav}{\yepv}
\label{eq_25}    
\end{equation}

where\\
$\yeav$ = Energy yield for AV farm  \\
$\yepv$ = Energy  yield for traditional standalone solar PV system at $\textit{p}/\textit{h}\approx2$ facing N/S at standard tilt.

\subsection{Effect of Temperature and soiling on PV energy}
The analysis for temperature effects for bifacial and monofacial PV schemes is shown in supplementary information (\cref{Supplementry}) and is incorporated in all results. 
To estimate the soiling effect for Lahore, recently published experimental data is used \cite{ullah2019investigation, imran2019effective} which reports the soiling rate of 0.9$\%$/day for the tilt angle of 20$^\circ$. The relative soiling loss for vertical tilt is reported to be negligible \cite{ullah2019investigation}. Based on this, the power loss due to soiling for mono-N/S PV is calculated to be 4-7$\%$ higher as compared to \textit{bi-E/W} assuming the cleaning schedules of 1 week and 2 weeks, respectively. Although we emphasize the superior soiling performance for the vertical panels, soiling loss is not specifically included in the PV output shown in Section III because of its variability associated with cleaning frequency and the seasonal rainfall patterns.

\section{Results and Discussion}
\label{results}

The performance of agrivoltaic system has been assessed for the \textit{N/S} tilted and \textit{E/W} vertical PV schemes for Lahore (31.5$^\circ$ N, 74.3$^\circ$ E) for a range of panel density including the standard ($\textit{p/h} = \text{2 }$), half ($\textit{p/h} = \text{4 }$) and double ($\textit{p/h} = \text{1 }$) across all months of the year. 

\subsection{PV energy generation and the transmitted PAR}

\begin{figure}[!b]
    \centering
    \includegraphics[width=0.8\linewidth]{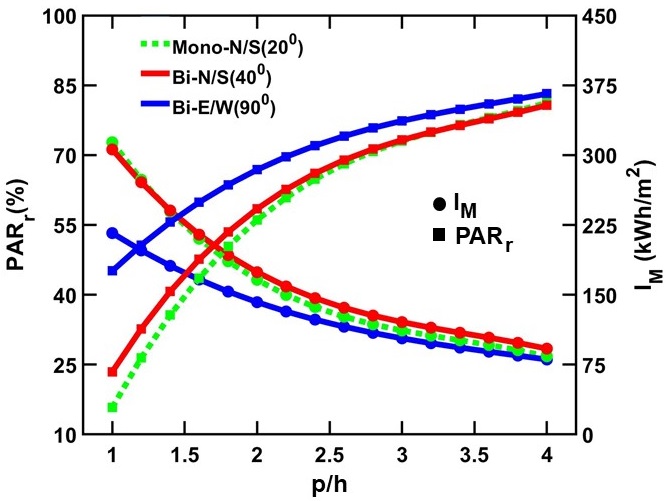} 
    \caption{$\ggr$ and $\textit{I}_\text{M}$ as a function of \textit{p/h} for E/W and N/S PV configurations.}
    \label{Fig_6}
\end{figure}

\begin{figure}[!t]
  \centering
  \subfigure
  {
  \includegraphics[width=0.75\linewidth]{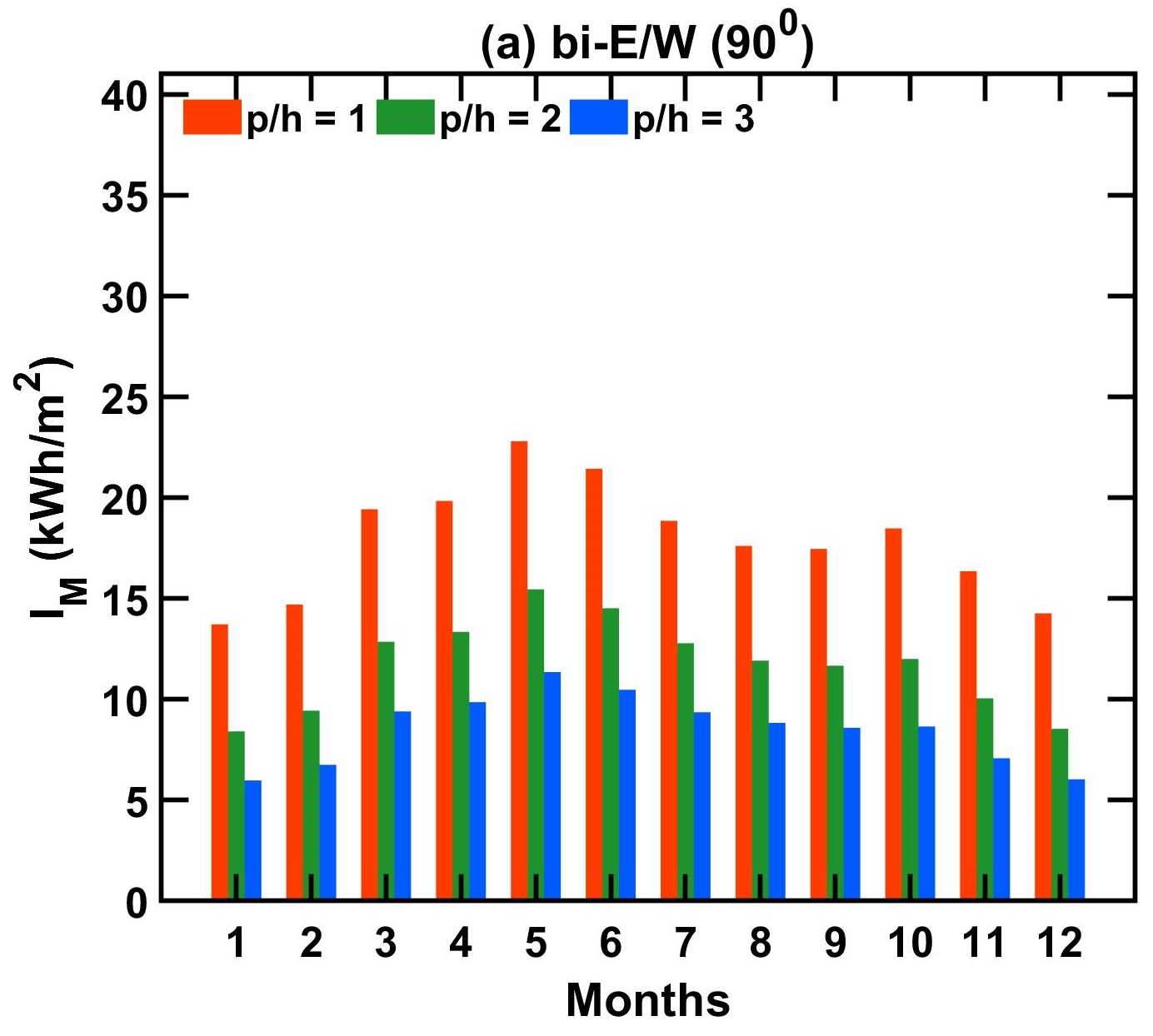}
  }
\subfigure
{
  \includegraphics[width=0.75\linewidth]{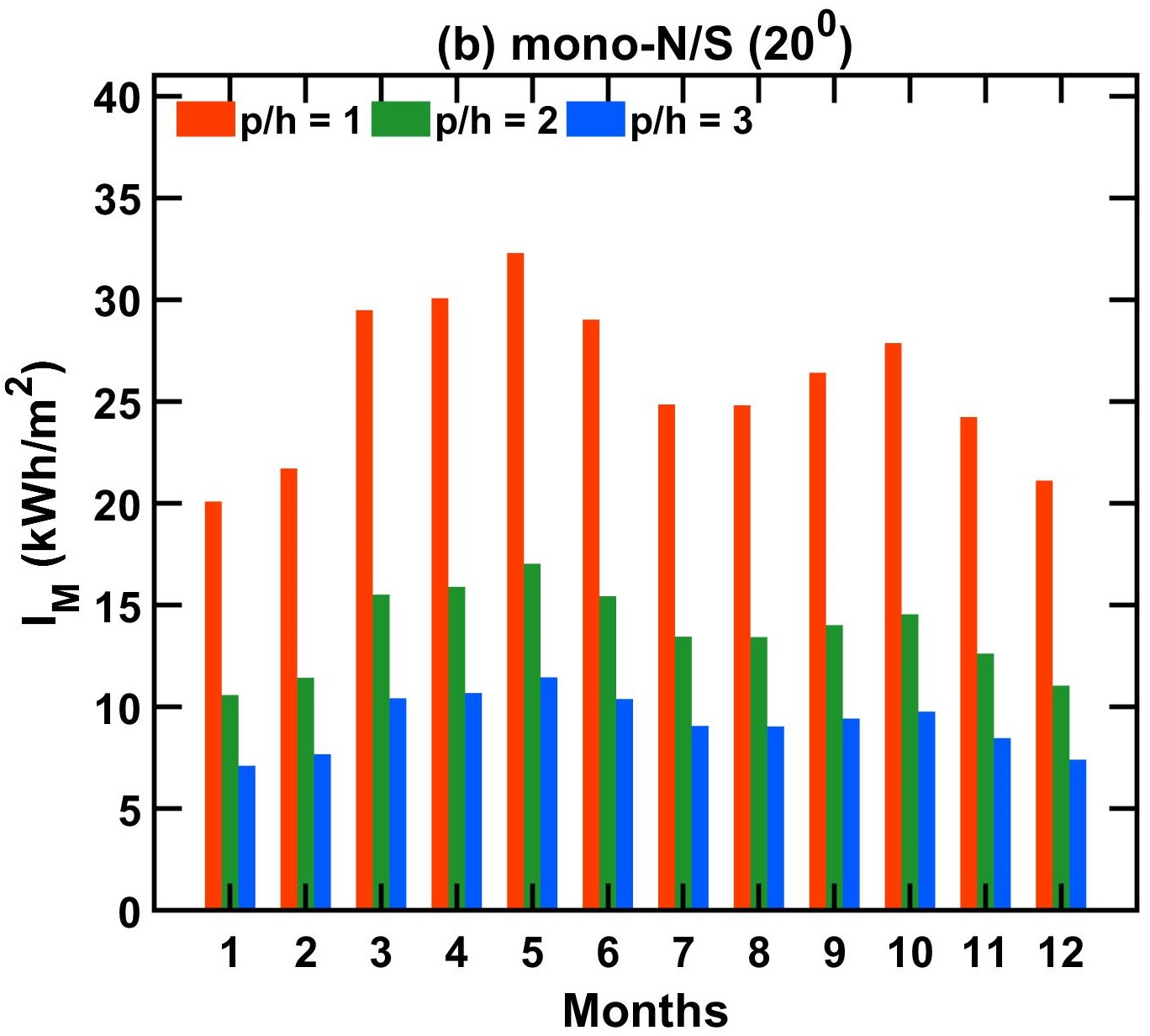}
  }
\caption{Energy generated in (a) \textit{bi-E/W} PV, and (b) \textit{mono- N/S} PV farm tilted at $\beta = \text{20}^\circ$ during all months of the year for different values of panel density.}
\label{Fig_7}
\end{figure}

\begin{figure}[!t]
  \centering
\subfigure
{
  \includegraphics[width=0.70\linewidth]{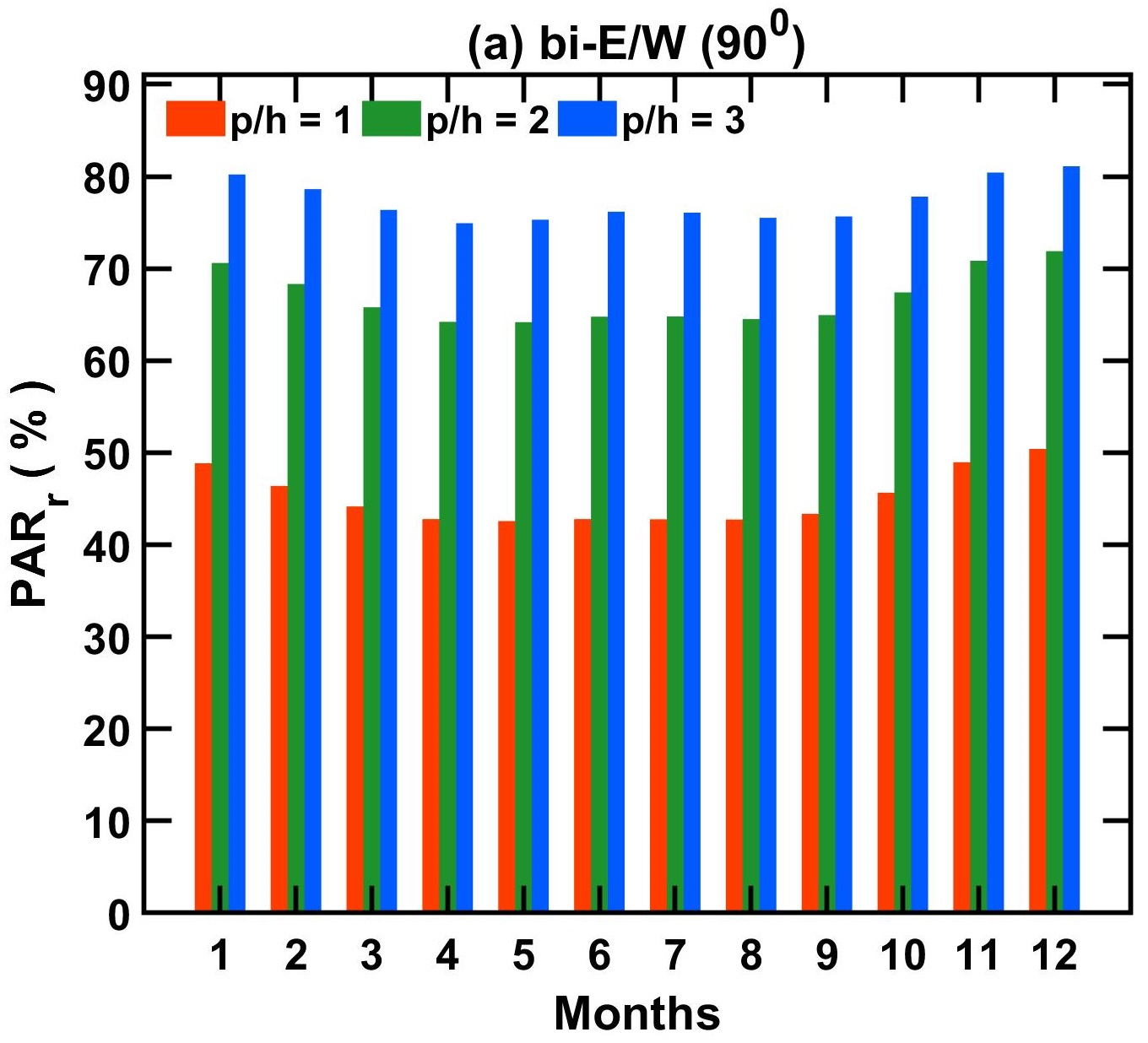}
  }
\subfigure
  {
  \includegraphics[width=0.70\linewidth]{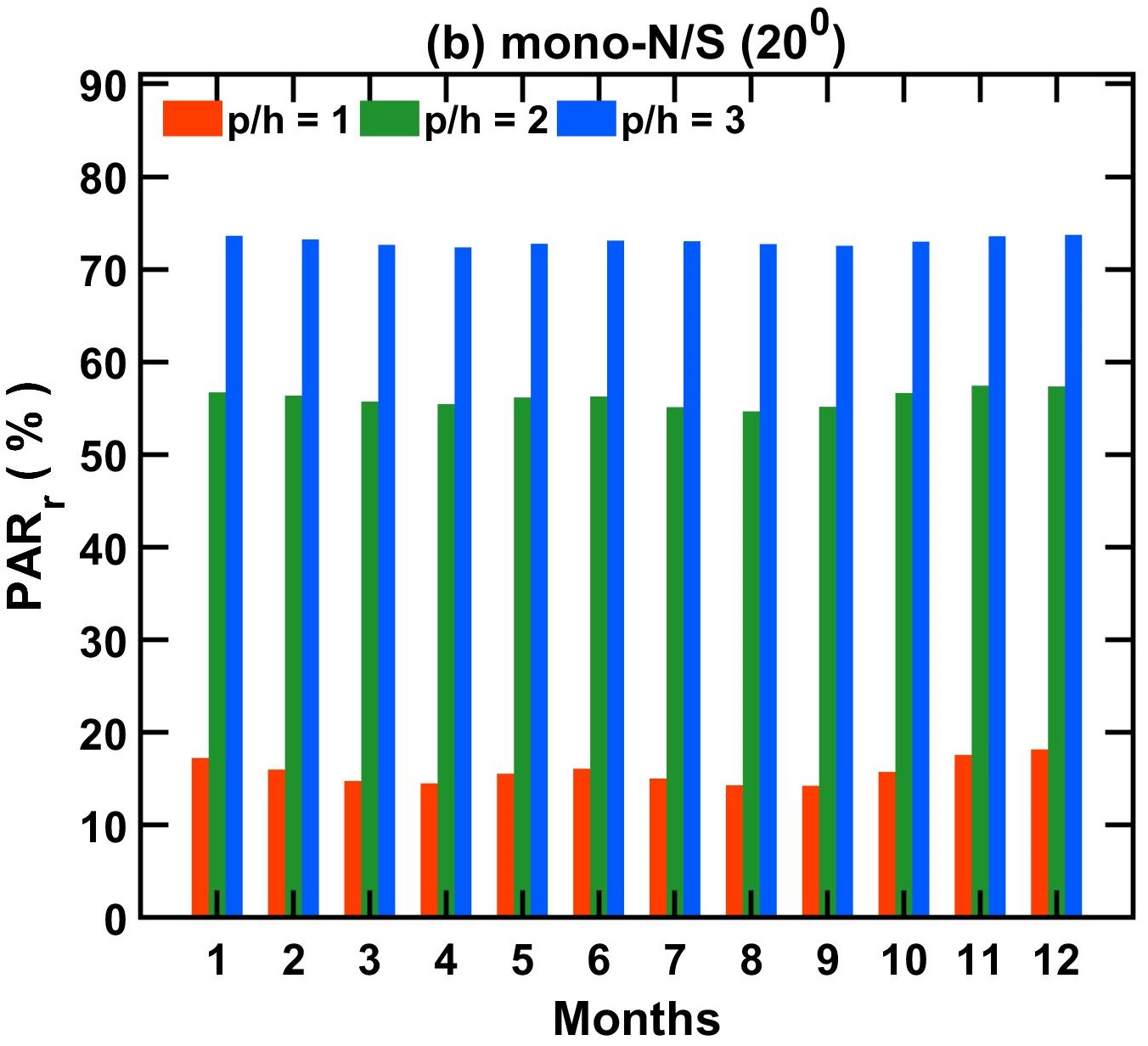}
  }
\caption{$PAR_r$ for (a) \textit{bi-E/W} PV, and (b) \textit{mono-N/S} PV farm tilted at $\beta = \text{20}^\circ$ during all months of the year for different values of panel density.}
\label{Fig_8}
\end{figure}

\cref{Fig_6} illustrates the annual PV energy production and $PAR_r$ for the PV schemes under consideration. For low density of panels (p/h$\approx$4), mutual shading between the panels is negligible while the energy production and $PAR_r$ for the PV schemes are relatively similar. As the panel density is increased, a couple of effects become important due to which the PV output and $PAR_r$ for \textit{bi-E/W} significantly diverge as compared to the N/S tilted schemes. Firstly, the mutual shading between the panels starts to increase dominantly for bi-E/W which lowers its relative PV energy yield while simultaneously increasing the $PAR_r$. This effect is stronger for bi-E/W due to its orientation as well as its relatively higher (vertical) tilt angle. Secondly, the masking of PV panels for the diffused $PAR_r$ strongly reduces for \textit{N/S} schemes due to their smaller tilt angles. This is most prominent in mono-N/S due to its smallest tilt angle for which only 16\% of \textit{PAR} is transmitted to the ground at very high density of panels (\textit{p}/\textit{h} $\approx$ \text{1}). These trends illustrate a trade-off between PV energy and $PAR_r$. The monthly PV energy production and $PAR_r$ for the \textit{N/S} and \textit{E/W} faced schemes are shown in \cref{Fig_7} and \cref{Fig_8} respectively for varying panel density. The monthly PV energy trends for \textit{E/W} and \textit{N/S} schemes are similar and follow the expected variation in solar irradiation intensity across the months. The $PAR_r$ values for \textit{bi-E/W} PV are slightly higher relative to the \textit{mono-N/S} in general for all panel densities. This trend becomes more significant for winter months and for high panel density.

\subsubsection{Energy yields and Land Productivity Factor}

\begin{figure}[!t]
\centering
\subfigure
  {
  \includegraphics[width=0.68\linewidth]{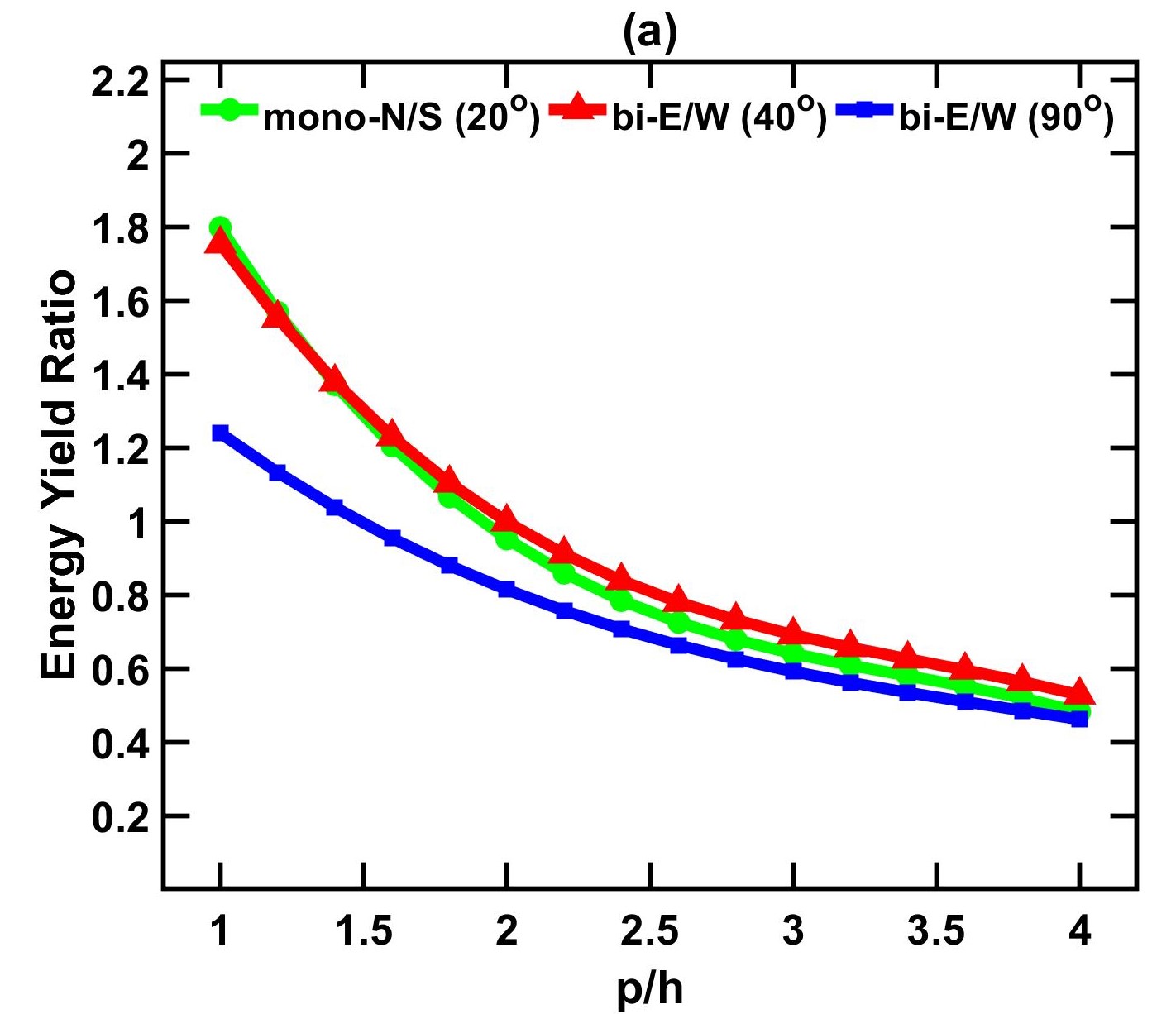}
  }
\subfigure
  {
  \includegraphics[width=0.68\linewidth]{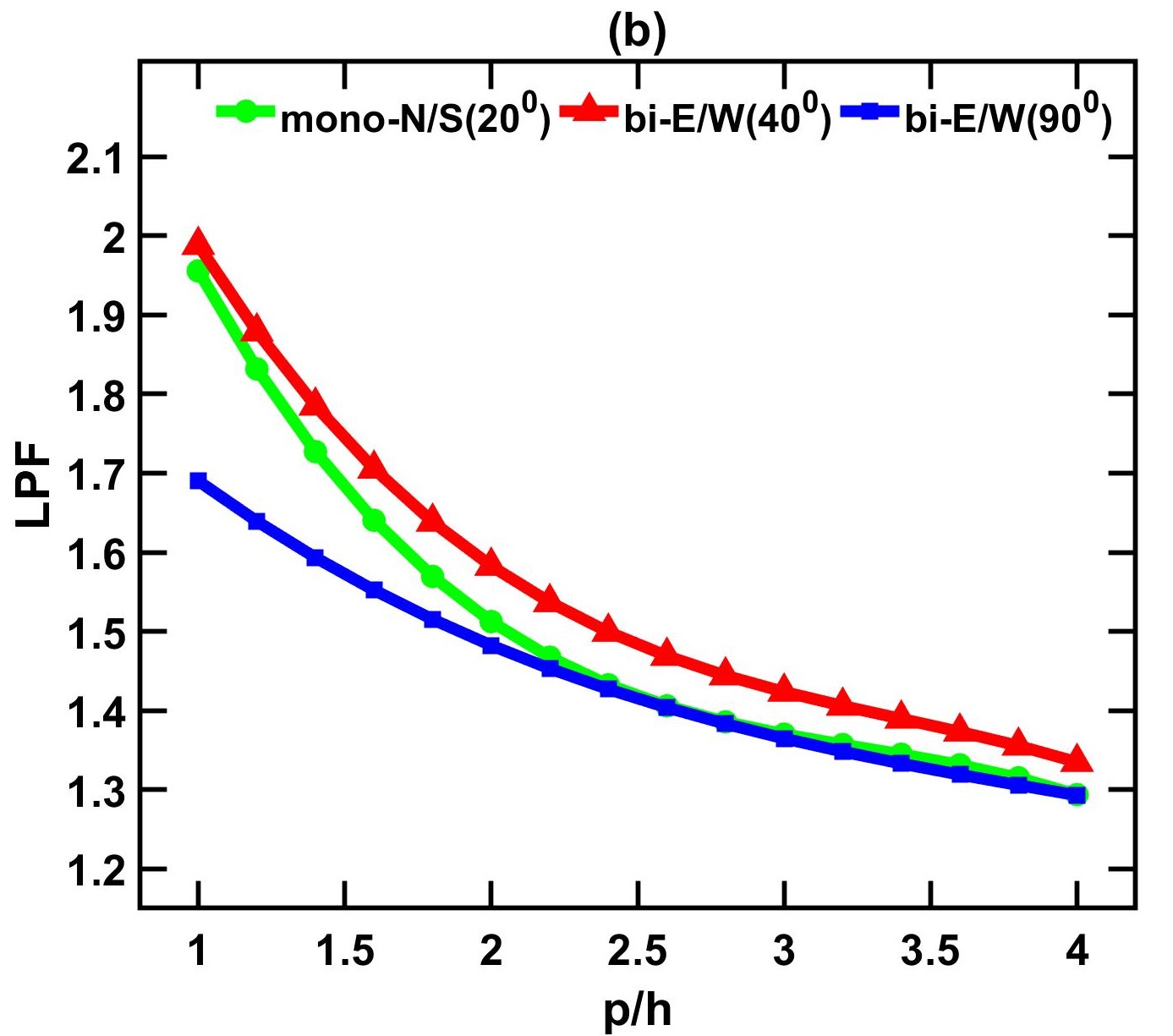}
  }
\caption{(a) Energy Yield Ratio, and (b) \textit{LPF} as a function of \textit{p/h} for (\textit{bi-E/W}) and (\textit{mono-N/S} and \textit{bi-N/S}) PV configurations.}
\label{Fig_12}
\end{figure}

The energy yield and \textit{LPF} (as defined in equation (\textit{16})) are shown in \cref{Fig_12} for \textit{bi-E/W} and \textit{mono (bi)-N/S} PV as a function of the panel density. The energy yield for \textit{bi-E/W} suffers due to mutual shading as \textit{p}/\textit{h} is decreased as already discussed which also results in a relatively higher $PAR_r$. At the half density (\textit{p}/\textit{h} $\approx$ \text{4}), \textit{LPF} is $\sim$1.35 and is not significantly different between \textit{N/S} and \textit{E/W} faced schemes. The \textit{LPF} however drops for \textit{bi-E/W} as the panel density is increased above the half density. At full density, \textit{LPF} ranges around 1.5-1.6 with \textit{N/S} faced panels providing slightly higher productivity.

\begin{figure}[!t]
  \centering
  \includegraphics[width=0.75\linewidth]{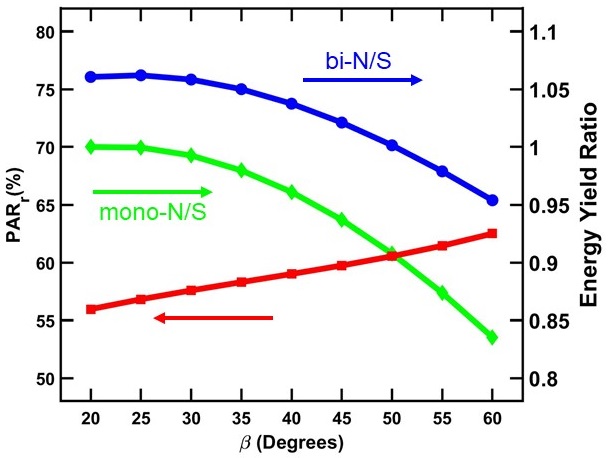}
\caption{Energy Yield Ratio and $PAR_r$ as a function of tilt angle.}
\label{Fig_16}
\end{figure}

\subsection{Effect of Tilt Angle}
For fixed tilt \textit{N/S} faced solar panels PV arrays at locations in the northern hemisphere, the PV tilt angle that is optimized for annual energy yield is known to be somewhat closer to the latitude of the location \cite{ullah2019investigation}\cite{kaddoura2016estimation}\cite{khorasanizadeh2014establishing} whereas the monthly optimal PV tilt angle for energy yield is known to decrease from winter to summer months. For AV systems, the optimal tilt angle has to be adjusted according to the desired trade-off for the sunlight sharing between panels and crops.
\cref{Fig_16} shows that the annual energy yield drops by 10$\%$ to 16$\%$ as the tilt angle varies from 20$^\circ$ to 60$^\circ$ for mono and bi-N/S PV schemes respectively at \textit{p/h=2}. The annual $PAR_r$, on the other hand, shows a respective increase of ~13\% for the same variation in the N/S tilt angle. This shows that the tilt angle adjustment can be a useful knob to fine tune the required balance of PV/$PAR_r$ once the system has been installed at a certain array density. 

\begin{table}
\caption{Summary of optimal array design for bi-E/W and mono(bi)-N/S PV configurations ensuring $\ge$80\% $PAR_r$ yield.}
\label{Table_New1}
\centering
\renewcommand{\arraystretch}{1.3}
\renewcommand{\tabcolsep}{0.8mm}
\setlength{\arrayrulewidth}{0.8pt}
\arrayrulecolor{white}
\begin{tabular}
{
M{6em}!{\color{white}\vline}
M{9em}!{\color{white}\vline}
M{6em}!{\color{white}\vline}
M{5em}
}

\rowcolor{blue!40}  PV Scheme &	$\textit{p}/\textit{h}$, (80\% $\textit{PAR}_\text{r}$ Yield) & PV Energy Yield &	LPF \\ \hline

\rowcolor{blue!30}  mono-N/S & 3.9 & 0.50 & 1.3 \\

\rowcolor{blue!20}  bi-N/S & 3.9 & 0.55 & 1.35 \\

\rowcolor{blue!30}  bi-E/W & 3.5 & 0.53 & 1.33 \\

\end{tabular}
\end{table}

\begin{table}
\caption{Summary of optimal array design for bi-E/W and mono(bi)-N/S PV configurations ensuring 80\% energy yield.}
\label{Table_New}
\centering
\renewcommand{\arraystretch}{1.3}
\renewcommand{\tabcolsep}{0.8mm}
\setlength{\arrayrulewidth}{0.8pt}
\arrayrulecolor{white}
\begin{tabular}
{
M{6em}!{\color{white}\vline}
M{9em}!{\color{white}\vline}
M{6em}!{\color{white}\vline}
M{5em}
}

\rowcolor{blue!40}  PV Scheme &	$\textit{p}/\textit{h}$, (80\% PV Energy Yield) &	$\textit{PAR}_\text{r}$ Yield &	LPF \\ \hline

\rowcolor{blue!30} mono-N/S & 2.6 & 0.62 & 1.42 \\

\rowcolor{blue!20} bi-N/S & 2.5 & 0.63 & 1.43 \\

\rowcolor{blue!30} bi-E/W & 2.1 & 0.67 & 1.47 \\

\end{tabular}
\end{table}

\subsection{Optimal PV array density for given food-energy requirements}

The design of an optimal array density for AV depends on system’s food-energy requirements. As there is a trade-off between the relative food and energy production as the panel density is varied so an optimal array density is needed to provide the best balance between PV output and PAR reaching to the crops. Here we explore this for a couple of AV design scenarios (i) For design A, we optimize the array density ensuring that at least 80$\%$ of the \textit{PAR} is transmitted to the ground relative to the open sun condition, and, (ii) For design B, we optimize the array density that ensures 80$\%$ of the PV energy yield is obtained relative to the full density \textit{mono-N/S} PV system. \textit{Table I} and \textit{Table II} show the required PV density and the respective \textit{LPF} obtained for these designs respectively. \textit{Table 1} shows that the optimal PV array density to implement design A requires \textit{p}/\textit{h} $\approx$ \text{4} for all three schemes. On the other hand, the required PV array density to implement design B is close to \textit{p}/\textit{h} $\approx$ \text{2.6} and \textit{p}/\textit{h} $\approx$ \text{2} for \textit{N/S} and \textit{bi-E/W} PV schemes respectively. This relatively higher array density for design B results in a drop of $PAR_r$ to ~60-70$\%$.

\section{Summary and Conclusions}
\label{conclusions}
In this study, we explore the relative effectiveness of fixed tilt vertical bifacial \textit{E/W} faced PV array for the first time for agrivoltaics. 
We develop a rigorous analytical framework to precisely calculate the sunlight intercepted by elevated PV arrays and the transmitted PAR below the elevated PV array.
 We conclude that:

\renewcommand\thefigure{S1}
\begin{figure}[!t]
    \centering
    \includegraphics[width=0.65\linewidth]{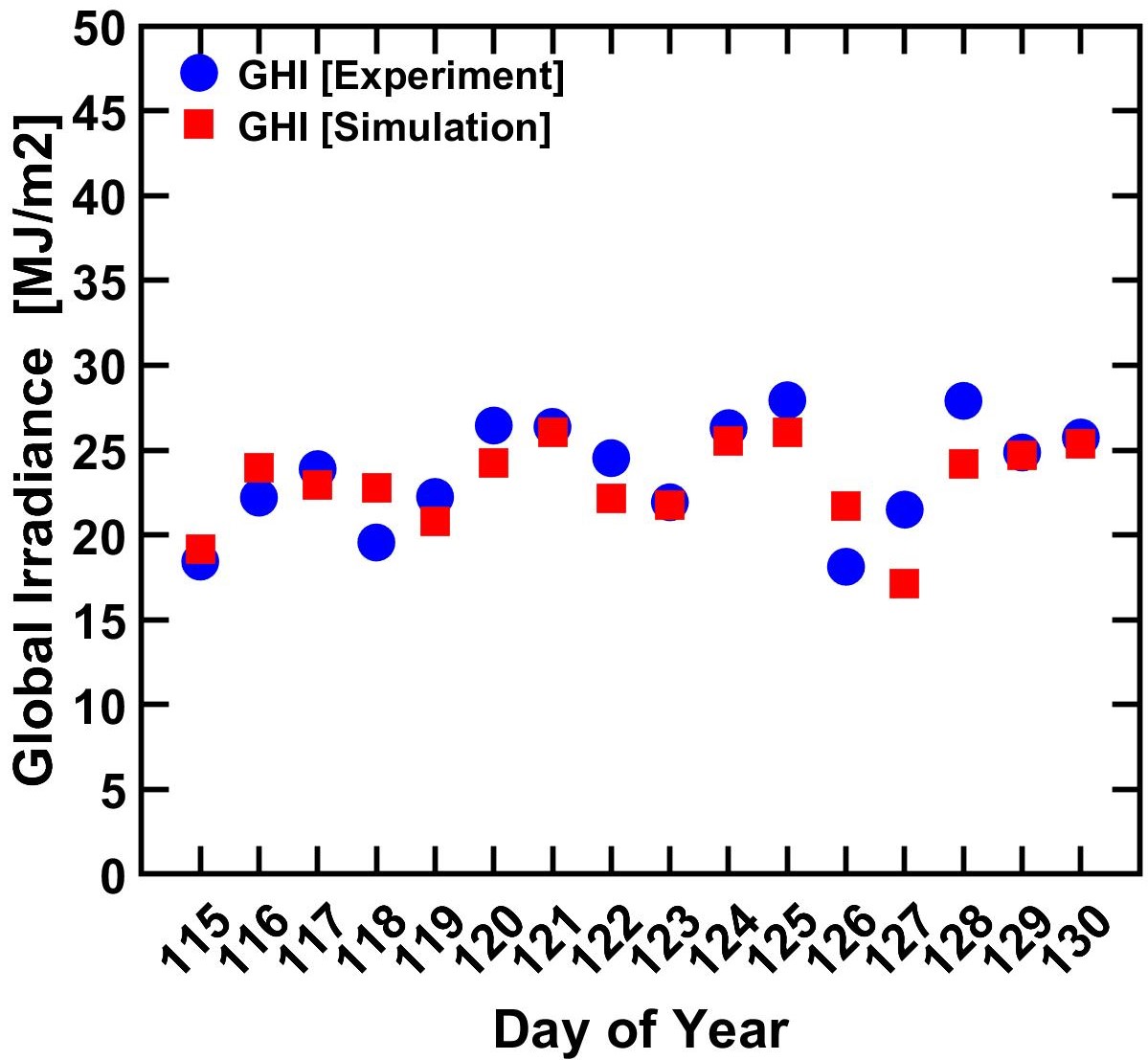}
    \caption{Global horizontal irradiation from the model output vs. measured field data \cite{marrou2013microclimate}}
    \label{Fig_S2}
\end{figure}

\begin{itemize}
\item Relative yields for PV energy and $PAR_r$ are similar for the vertical \textit{bi-E/W} PV and \textit{mono(bi)-N/S} PV when the panel density is half or lower relative to that of the standard ground mounted PV farms. For denser PV arrays , \textit{bi-E/W} results in a higher crop yield at the cost of reduced energy yield. The relative $PAR_r$ yield for \textit{bi-E/W} increases as compared to \textit{N/S} schemes when the array density is increased. 
 
\item For PV arrays closer to the full density, the annual $PAR_r$ drops to 55—65$\%$ while the relative PV yield for the vertical \textit{bi-E/W} drops to 80$\%$ with \textit{LPF} close to 1.5.

\item For half density PV arrays, the annual $PAR_r$ of ~80$\%$ is obtained while the relative annual PV yields for vertical \textit{bi-E/W} and \textit{mono(bi)-N/S} PV drops to 50–55\% with \textit{LPF} of 1.3–1.35.

\item For N/S tilted PV schemes, adjustments in the tilt angle can enable fine-tuning of the sunlight balance needed between PV and $PAR_r$. A modulation capability of 10-15$\%$ is calculated while varying the N/S tilt angle between 20$^\circ$ to 60$^\circ$. 

\item The land productivity estimates for \textit{mono(bi)-N/S\textit} tilted and \textit{bi-E/W} vertical \textit{E/W} faced panels are although not significantly different for PV arrays that have half ($\textit{p/h} = \text{4 }$) or lower density, some unique benefits for the \textit{bi-E/W} vertical PV including minimal PV soiling (hence better generation efficiency), minimal obstruction to rainfall and farm machinery, small land coverage and possibly lower elevation requirements can make vertical \textit{bi-E/W} AV farms preferable especially for hot and arid climates having significant PV soiling concerns.
\renewcommand\thefigure{S2}
\begin{figure}[!b]
\centering
\subfigure
  {
  \includegraphics[width=0.60\linewidth]{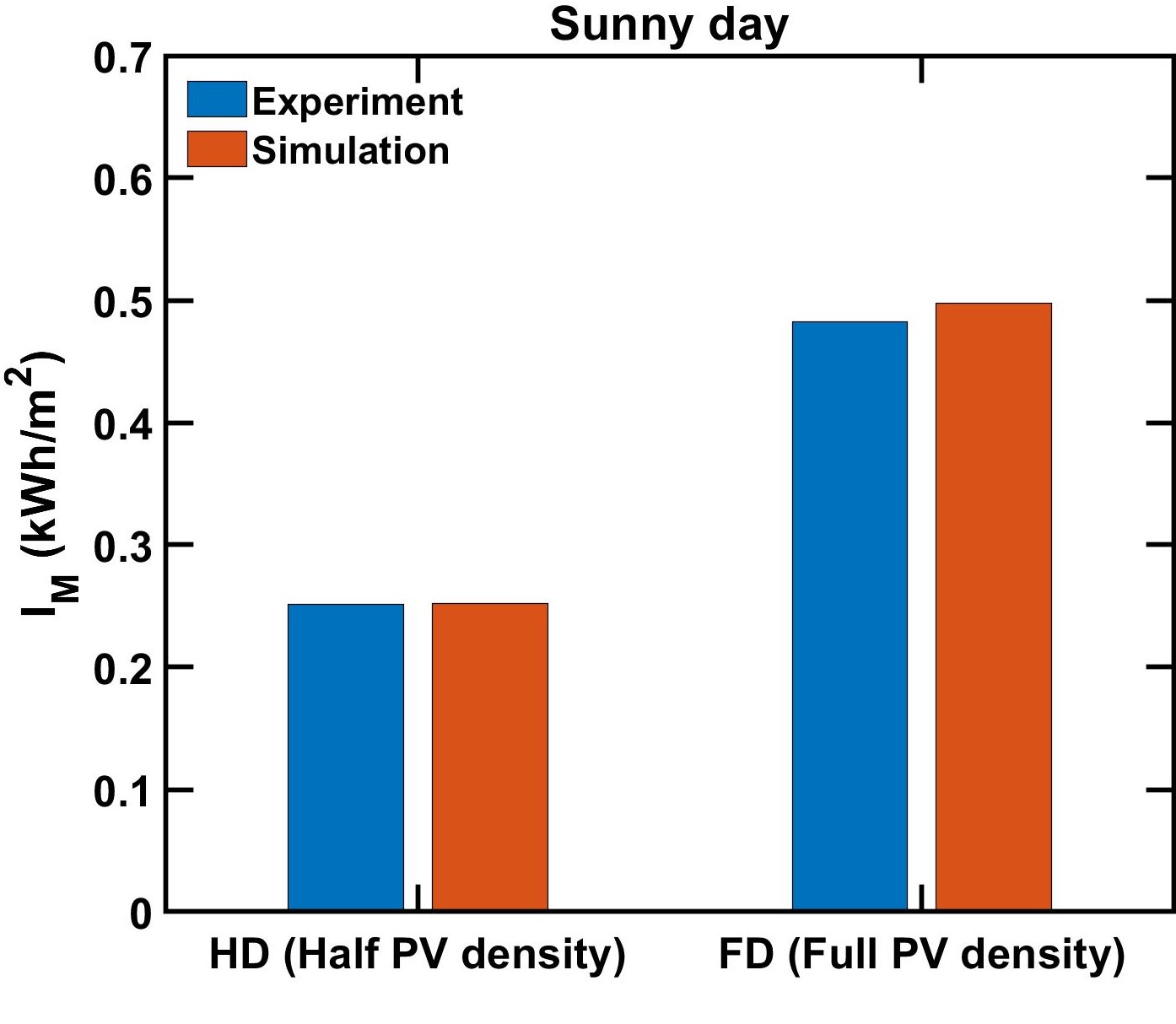}
  }
\subfigure
{
  \includegraphics[width=0.60\linewidth]{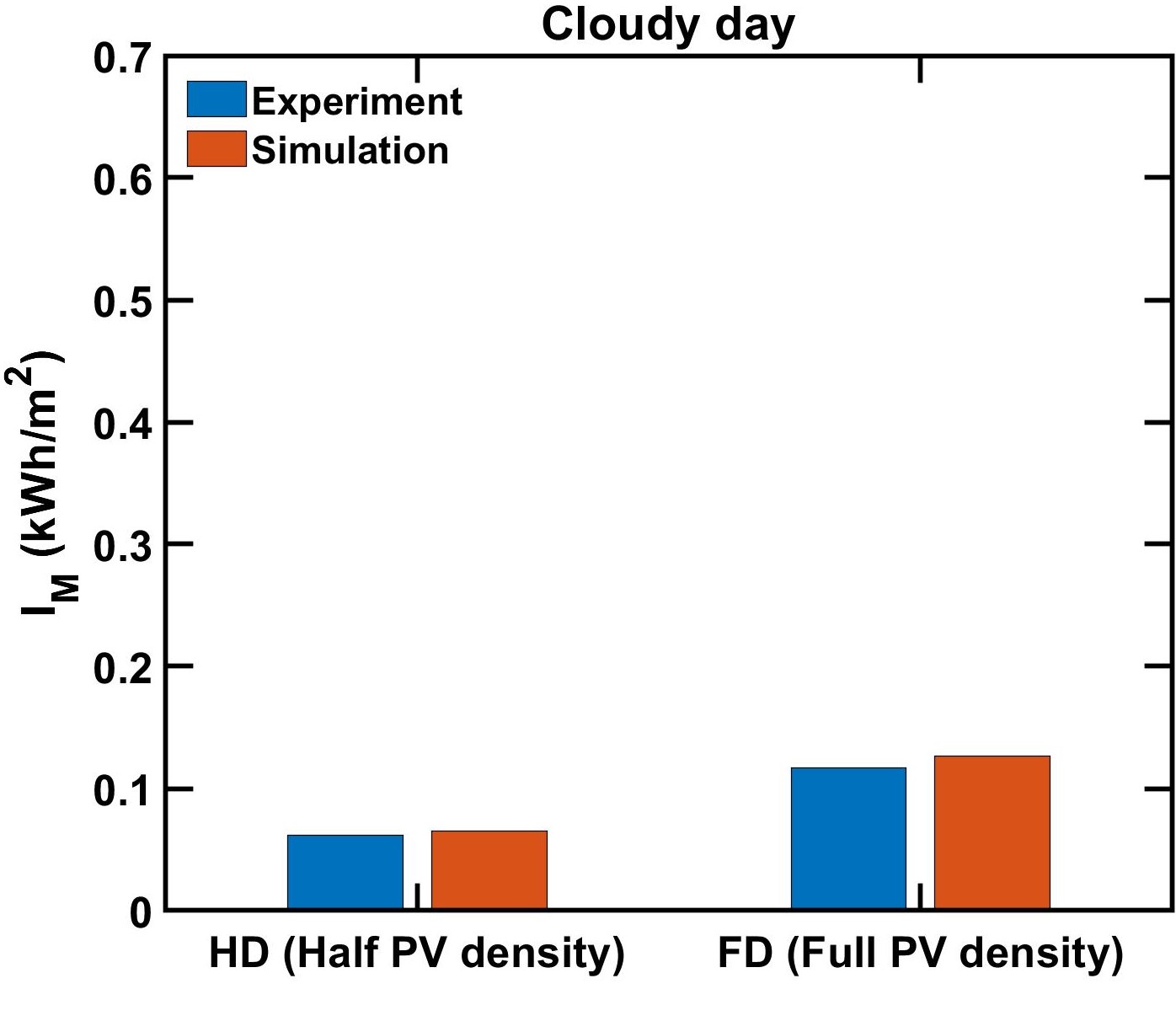}
  }
\caption{PV energy output comparison between our model and measured field data \cite{valle2017increasing} for a sunny and cloudy day.}
\label{Fig_S3}
\end{figure}

\item Finally, the economic factors which are deemed beyond the scope for this study should be considered to determine the ultimate choice of the preferable PV scheme for a particular AV application. In particular, since the bifacial solar panels can be somewhat more expensive, the capital cost for the bi-facial system could be relatively higher. The vertical panels, on the other hand, could have a significantly lower cleaning requirement particularly for soiling intense environments. The relative levelized cost of energy (\textit{LCOE}) should therefore be carefully evaluated considering the trade-off between a higher capital cost vs. low cost for periodic cleaning.  
\end{itemize}

\section*{Supplementary Information}
\label{Supplementry}

\renewcommand\thefigure{S3}
\begin{figure}[!b]
 \centering
\subfigure
 {
  \includegraphics[width=0.60\linewidth]{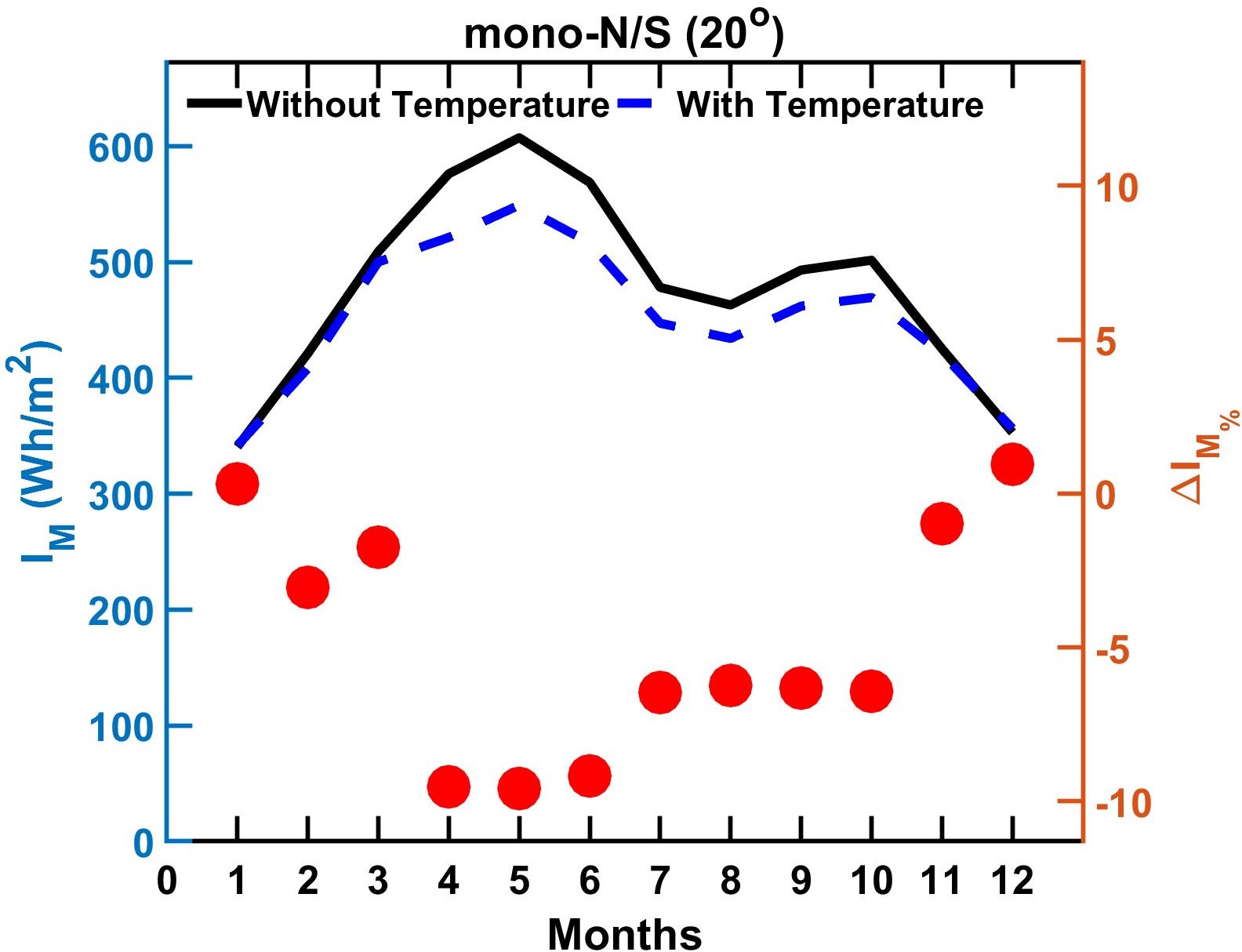}
  }
\subfigure
{
  \includegraphics[width=0.60\linewidth]{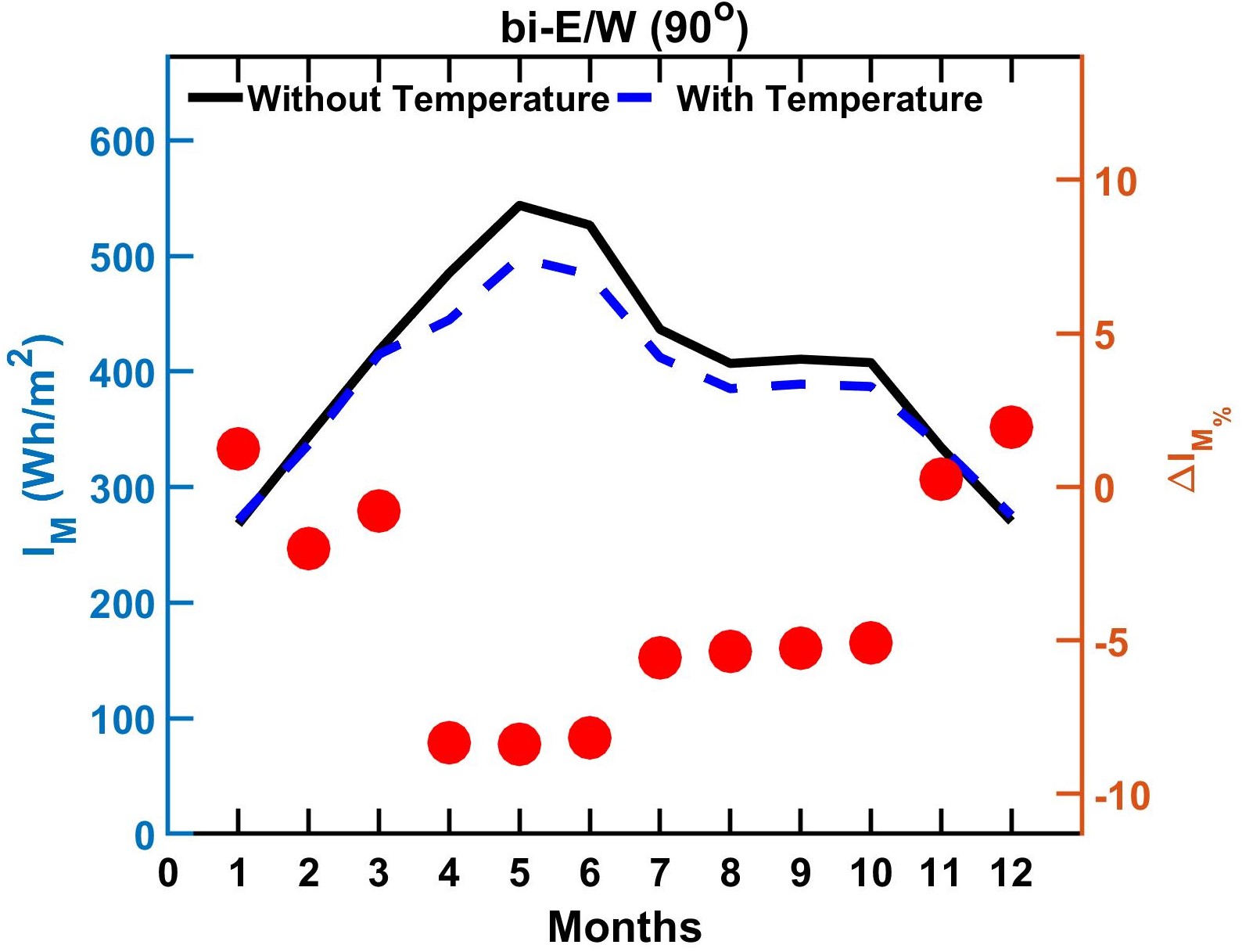} 
  }
\caption{PV energy output for each month with and without the temperature effect for (left) mono-N/S and (right) bi-E/W PV schemes. The \% difference in the PV output due to the temperature effect is also shown (circles).}
\label{Fig_S5}
\end{figure}
\subsection*{Model validation}
The global horizontal irradiation (\textit{GHI}) and the PV energy output calculated using our model are compared with the experimental measurements in a reported AV field study at Montpellier, France \cite{marrou2013microclimate}. \textit{GHI} measurement data across the range of 16 days and PV energy output data for a sunny and cloudy day match well with the model output for the same location as shown in \cref{Fig_S2} and \cref{Fig_S3} respectively implying a good validity for our model.

\subsection*{Effect of temperature on PV energy output}

An empirical model developed by King \textit{et al.,} \cite{king2004photovoltaic}  at Sandia National lab has been used to quantify the temperature effect for PV energy output. The module temperature is related to the ambient temperature ($T_a$), incident irradiation (\textit{E}), and wind speed (\textit{WS}) by \cite{king2004photovoltaic} : 
\begin{equation}
T_m = E.e^{a+b. W_S}+ T_a
\label{eq_s1}    
\end{equation}
where a and b are the empirical constants. The temperature corrected efficiency $\eta_T$ of the PV cell is then related to the standard cell efficiency $\eta_{STC}$:
\begin{equation}
\eta_T = \eta_{STC} [1 - T_c \Delta T]
\label{eq_s2}    
\end{equation}

where ($T_C$) is the temperature coefficient (assumed to be 0.41$\%$ /$^\circ$C \cite{pveducation} for crystalline silicon panels) and $\Delta$T is the difference in the ambient and cell temperatures. 

\bibliographystyle{IEEEtran}
\bibliography{mybib}

\end{document}